\documentclass[aps,amssymb,amsmath,prl,twocolumn,superscriptaddress]{revtex4}

\usepackage{graphicx}
\usepackage{amssymb}
\usepackage{epstopdf}
\def\be{\begin{equation}}   \def\ee{\end{equation}}
\def\eq#1{{Eq.(\ref{#1})}}    \def\fig#1{{Fig.\ref{#1}}}

\def\kT{k_{{}_{\rm B}}T}

\begin{document}

\title{Force balance and membrane shedding at the Red Blood Cell surface}
\author{Pierre Sens}
\affiliation{Physico-Chimie Th\'eorique (CNRS UMR 7083)\\ ESPCI, 10 rue Vauquelin, 75231 Paris Cedex 05 - France\\ pierre;sens@espci.fr}
\author{Nir Gov}
\affiliation{Dept. of Chemical Physics - The Weizmann institute of Science\\
P.O.B. 26, Rehovot, Israel 76100 \\ nirgov@wisemail.weizmann.ac.il}

\date{\today}



\begin{abstract}
During the aging of the red-blood cell, or under conditions of
extreme echinocytosis, membrane is shed from the cell plasma
membrane in the form of nano-vesicles. We propose that this process
is the result of the self-adaptation of the membrane surface area to the elastic stress imposed by the spectrin cytoskeleton, via the local buckling of membrane
under increasing cytoskeleton stiffness. This model introduces the
concept of force balance as a regulatory process at the cell
membrane, and quantitatively reproduces the rate of area loss in
aging red-blood cells.
\end{abstract}

\maketitle

The membrane of red-blood cells (RBCs) is a fluid sheet of lipids and proteins, connected through node complexes to a two-dimensional cytoskeleton primarily composed of flexible spectrin protein filaments  \cite{bennett}. The mechanical properties and the shape of the RBC result from a competition between the bending elasticity of the cell membrane and the shear and dilation elasticity of the cytoskeleton \cite{evans,wortis}.  Because of the relative structural simplicity of the spectrin network, the RBC is a system of choice to study the co-organization of the lipid bilayer and the cytoskeleton cortex at the plasma membrane  \cite{evans}.

Over the natural life-span of the RBC,  the cell looses a large fraction of its membrane ($\sim20\%$ over $120$ days in humans and $\sim10\%$
over $50$ days in rabbits \cite{mohandas1992,bosman2004}) by the shedding of small vesicles containing hemoglobin but devoid of cytoskeleton  \cite{bosman2004,palek1977,backman1998}. The  shed vesicles have a size (100-200nm  \cite{bennett}) similar to the mesh-size of the underlying spectrin network  \cite{bosman2004,palek1977,salzer2002}, and may take the form of elongated cylinders  \cite{iglic2004}. The shedding of vesicles is also observed under conditions that trigger the transition into echinocyte shape, such as ATP depletion and Ca$^{2+}$ loading  \cite{bucki}. In this case, the shedding of cytoskeleton-free vesicles often occurs at the tip of membrane spicules containing cytoskeleton  \cite{iglic2004,palek1989,chien1987}. The shedding of membrane area is crucial to the fate of the cell, as a decreasing surface-to-volume ratio is thought to be connected to the phagocytosis of old RBCs. It is this process of membrane shedding that we model in this work.

\begin{figure}[h]
\centerline{\includegraphics[width=8.5cm]{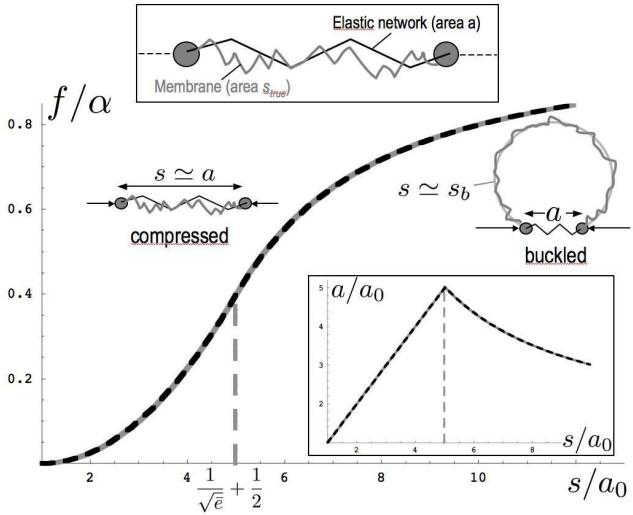}}
\caption{\label{unit_buck} \small Energy of a patch of fluctuating membrane of apparent area $s$ fit to an elastic frame of rest area $a_0$. The energy shows an inflexion point when the membrane buckles. The gray line is the numerical optimization of \eq{en1} with \eq{entot} and the dashed line is the asymptotes \eq{f-b}, for $\alpha=500$ ($\kappa=20\kT$) and $\bar e=5\ 10^{-2}$ ($e=10^{-5}J/m^2$ and $a_0=(100nm)^2$). Inset: The deformation of the cytoskeleton (area $a$) is maximal at the transition.}
\end{figure}

When the entire membrane of the RBC is removed under physiological
conditions, the cytoskeleton shrinks to a 3 to 5-fold smaller area  \cite{atprafi}. This indicates that the
cytoskeleton is stretched by its attachment to the cell
membrane \cite{wortis,discherstress}, and therefore exerts a
compression force on the membrane. 
As the cell ages, its observed
stiffness increases by $\sim20\%$ \cite{fricke,sutera}, as does the
density of its cytoskeleton network by about $\sim30-40\%$
 \cite{ostafin}. A recent model  \cite{nirsam} proposes that these
changes may be related to the diminishing levels of ATP as the cell
ages  \cite{tochner}. This model relates the ATP level to the degree
of dissociations in the cytoskeleton network; Lower ATP levels
correspond to a stronger cytoskeleton network, and results in larger
compressive forces on the cell membrane. These forces are balanced
by the membrane bending stress  \cite{wortis}, which may either  be
of entropic origin (from constraints on the membrane thermal fluctuations), or resulting from an overall membrane  curvature
(blebbing). We show below that beyond a compression threshold, the
membrane undergoes a buckling transition at the scale of the
spectrin mesh-size, and one or a few membrane blebs grow in size in
order to release the overall stress. We first analyze the
buckling transition of a single membrane-cytoskeleton unit. The
thermally fluctuating membrane is assumed flat at large lengthscales
(we do not account for the global shape of the RBC), while the
spectrin filaments are treated as phantom Hookean springs
 \cite{fournier}, attached at their ends to the membrane
(\fig{unit_buck}). We then consider a large elastic network attached
to a membrane, and show that the lowest energy state beyond a
critical compression strain consists of an essentially stress-free
membrane connected to a large bleb which incorporates all the excess
membrane area. We argue that such a state is never reached because
the bleb pinches off the membrane to form a free vesicle before
reaching large sizes. Finally, we relate our results to the observed
shedding of vesicles during RBC aging.

Let's assume that one cytoskeleton unit, of area $a$, is connected
to a patch of membrane of area $s_{true}$ (\fig{unit_buck}), and
constrains its fluctuations  \cite{jbf_rbc}. There exist an optimal
unit area that leaves the fluctuations unconstrained. This area,
which we call $s$, is the apparent area of a freely fluctuating membrane and is
slightly smaller than the true area by a factor
$s/s_{true}=(1-1/\alpha\log{s_{true}/l^2})$ \cite{helfrich_servus}.
Here, $\alpha\equiv8\pi\kappa/\kT$ is the bending energy of a closed
sphere (compared to the thermal energy), and $l$ a small size of
order the membrane thickness. If the membrane is stretched ($a>s$),
or mildly compressed, it remains flat on average and its free energy
is \cite{helfrich_servus,sens_safran} 
\be
f_{fl}[s,a]=\frac{\pi}{2}k_BT\left(e^{\alpha
(a/s-1)}-1-\alpha(a/s-1)\right) 
\label{fm} 
\ee 
This energy is
quadratic ($\sim \pi/2 \alpha^2(a/s-1)^2$) for small compression
($\alpha(s/a-1)<1$), and becomes linear ($\simeq\pi/2
\alpha(1-a/s)$) for large compression. The energy of the flat
compressed patch eventually becomes larger than the bending energy
of a spherical cap, leading to the breaking of the up/down
symmetry and the formation of a curved bleb. We investigate the
buckling transition by assuming that a buckled membrane fluctuates
around a mean shape defined by a spherical cap of area $s_b\ge a$,
to be optimized (Fig.1). The energy of a buckled patch reads
\be
f_m[s,a]=8\pi\kappa\left(1-\frac{a}{s_b}\right)+f_{fl}[s,s_b]
\label{entot} 
\ee 
The first term is the bending energy of a
spherical cap of area $s_b$ on a frame of area $a$, and the second
the energy of the thermal fluctuations (\eq{fm}), assumed unaffected
by the curvature of the bleb. The membrane is flat on average if
$s_b=a$ and is buckled if $s_b>a$. Comparing \eq{fm} and \eq{entot}
shows that in the relevant limit ($\alpha\gg1$), the
optimal buckled area is $s_b\simeq s-2a/\pi\alpha$, and buckling is
to be expected for $s/a>1+2/\pi\alpha$.

In addition to the membrane entropic elasticity $f_m$, the total
energy of a membrane unit has a contribution from the cytoskeleton
elasticity, treated here as a simple spring-like quadratic energy:
 \be f=\frac{1}{2}ea_0\left(\frac{a}{a_0}-1\right)^2+f_m[s,a]
 \label{en1} \ee
Here, $e$ and $a_0$ are the stretching modulus and rest area of the
cytoskeleton unit mesh. The full free energy for one patch (\eq{en1}, with
\eq{entot}) is to be optimized with respect to two areas: the buckled membrane area $s_b$, and the area of the stretched cytoskeleton mesh ($a\geq a_0$). The buckling transition is controlled by the area mismatch
$s/a_0$ and the  ratio of the typical cytoskeleton to
membrane elastic parameters $\bar e\equiv e a_0/\alpha\kT$. In the
case of the RBC, we expect $\bar e\ll1$ and the buckling transition
is second order (the cytoskeleton area is continuous through the
transition, \fig{unit_buck}-inset). The numerical minimization of the full energy in shown
in \fig{unit_buck}. The buckling transition occurs for a critical
mismatch $s/a_0\simeq 1/\sqrt{\bar e}$ (for $\alpha\gg1$ and $\bar
e\ll1$), and the optimal meshsize and energy below and above the
transition (subscripts ``$f$'' and ``b'', respectively) are:
\begin{eqnarray}
a_f=s\quad;\quad \frac{f_f}{\alpha}=\frac{1}{2}\bar e\left(\frac{s}{a_0}-1\right)^2\cr
a_b=a_0\left(1+\frac{a_0}{\bar e s}\right)\quad;\quad \frac{f_b}{\alpha}=1-\frac{a_0}{s}-\frac{a_0^2}{2\bar e s^2}
\label{f-b}
\end{eqnarray}
Note that the area of the stretched cytoskeleton mesh increases with
increasing membrane area $s$ below the buckling transition, but
decreases with increasing $s$ above the transition (Inset of  \fig{unit_buck}).

We now proceed to describe the buckling instability of many connected patches. Under low compression, the membrane of the cell is equally shared by all cytoskeleton units, and if the membrane were not fluid, an increase of the cytoskeleton compressive stress would eventually lead to the buckling of every single membrane patch. Biological membranes are fluid, so once a unit buckles, it may grow into a large bleb
by feeding from the excess area of the surrounding units. This
reduces their energy without much increase of the energy of the bud, which saturates to the energy of a full sphere $\alpha$
for large area (\eq{f-b}). Mathematically, coexistence between flat and buckled units is allowed by the existence of an inflexion point in the energy of one isolated membrane patch (\fig{unit_buck})).

\begin{figure}
\centerline {\includegraphics[width=9 cm]{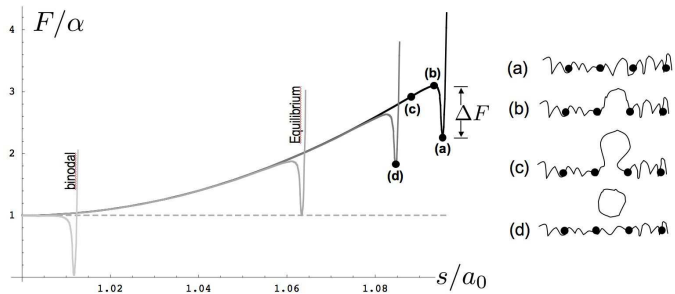}}
\caption{\label{full_buck} \small Energy of a collection of $N$ membrane + cytoskeleton units as a function of the membrane area per unit $s$ (from \eq{ftot}). Darker shades of gray corresponds to larger total membrane area $S$. The energy shows a narrow local minimum for an homogeneous flat membrane at $s=S/N$. A local minimum at $s\simeq a_0$ appears beyond the binodal point. It corresponds to stress-free units coexisting with a large membrane bleb. The homogeneous membrane {\bf (a)} is globally unstable for $S>S_{|_{eq}}$. After a waiting time depending on the nearness to the spinodal line (\eq{spineq}) a small bleb forms {\bf (b)}, grows  {\bf (c)}, and stochastically pinches off the cell {\bf (d)} reducing the total membrane area and leaving the system in a deeper local minimum until the next blebbing event. $N=10^4$, $\bar e=5\ 10^{-2}$ and $S/Na_0=1.01$ (binodal), $1.063$ (equilibrium), $1.085$, and $1.095$. The spinodal point ($S/Na_0=4.98$) is not shown.}
\end{figure}

Let's consider $N$ membrane-cytoskeleton units similar to the one shown in \fig{unit_buck}, with total membrane area $S$. We investigate a membrane composed of $(N-1)$ units of area $s$ that are in the flat (unbuckled) state, plus one unit of area $S-(N-1)s$, that may be buckled if such shape lowers the total energy:
\be
F^{tot}_{f,b}=(N-1)f_f\left(s\right)+f_{f,b}\left(S-(N-1)s\right)
\label{ftot}
\ee
where the energies in the flat and buckled state ($f_f$ and $f_b$ respectively) are given in \eq{f-b}. The parameter that controls the state of the membrane is the dimensionless average area per unit (area mismatch) $S/Na_0$, and the state of the membrane is unambiguously given by the area of the flat units $s$.
For small area mismatch, the only equilibrium state corresponds to a uniform flat membrane  with $s=S/N$. Upon increase of the mismatch, the system eventually reaches the binodal point. A local minimum of the total energy $F_{tot}$ appears for $s<S/N$, corresponding to an
inhomogeneous membrane state with one buckled unit (\fig{full_buck}).
For large area mismatch, the optimal conformation is such that all the flat units
are near their optimal area ($s\gtrsim a_0$), with all the excess area concentrated in the one nearly spherical bleb of energy
$F^{tot}_b\simeq\alpha$.

The coexistence between flat and buckled states requires the
equality of tensions: $\partial_sf_f(s)=\partial_sf_b(S-(N-1) s)$ ($\rightarrow
\partial_s F^{tot}_b=0$). The buckled state is
the equilibrium state beyond a total area $S_{|_{eq}}$ for
which buckled and flat states have the same energy
$F^{tot}_b=F^{tot}_f=N\alpha\bar e/2(S_{|_{eq}}/(N a_0)-1)^2$. Although
it is thermodynamically favorable to form a large bleb devoid of
cytoskeleton for $S>S_{|_{eq}}$, this process requires an
activation energy $\Delta F$ (see \fig{full_buck}). The energy barrier is very large (of order
$\alpha\simeq 500\kT$) close to equilibrium, decreases with increasing area mismatch,
and vanishes at the ``spinodal'' point $S_{|_{spin}}$  for which every single unit is unstable
with respect to buckling. In the relevant limit
($\bar e\ll1$ and $N\bar e\gg1$), the equilibrium and spinodal
mismatches are given by:
\be
\frac{S}{N a_0}|_{eq}\simeq
1+\sqrt\frac{2}{\bar e N} \quad;\quad \frac{S}{Na_0}|_{spin}\simeq
\frac{1}{\sqrt{\bar e}}
\label{spineq}
\ee

After the buckling of one membrane unit (\fig{full_buck}b), the system then evolves downhill (\fig{full_buck}c) toward
its global equilibrium state, which corresponds to a large bleb in coexistence with stress-free units ($s\simeq a_0$). The ``equilibrium bleb''  takes all the
excess area, and should thus be very large (area  of order $a_0\sqrt{N/\bar e}$ if buckling
occurs at equilibrium and $a_0 N/\sqrt{\bar e}$ near the spinodal
curve).

Although such large blebs are observed in blebbing cells, they
generally involve the unbinding of large patches ($\sim \mu m^2$)
of membrane from the cytoskeleton { \cite{SheetzBleb} and don't bud
off the cell. In aging RBC, small spherical or cylindrical vesicles
of size comparable to the cytoskeleton meshsize ($\sim (100nm)^2$)
are being shed. This phenomenon could have a kinetic origin and be
explained by the following simple argument. Once a unit membrane has
buckled, it initially grows to form a hemispherical cap (of bending
energy $\alpha/2$), from which it may grow further into a shape
approaching either a fully formed sphere or a  cylinder. The bending
energy of the cylinder ($\sim \alpha(s_b/(16a)+3/8$) is smaller than
the energy of the sphere ($\sim \alpha(1-a/s_b)$), as long as
$s_b<8a$, so we argue that the buckled unit should initially grow as
a cylinder.

The growing bleb should undergo large fluctuations, owing to the
fact that the membrane is under small (or even negative) surface
tension. One may thus expect the high curvature neck connecting the
bleb to the cell membrane (\fig{full_buck}c) to eventually close and
the bleb to pinch-off before reaching a large size, in agreement
with the experimental observations. Vesicle scission is expected to
be much faster than the RBC aging process and could be promoted by
phase separation of lipids and proteins within the membrane of the
bleb \cite{PatriciaTubes}. Once shedding has occurred, the system
finds itself in a new (deeper) local energy minimum
(\fig{full_buck}d), corresponding to the reduced total membrane area
$S$.

In the context of the RBC, the cytoskeleton stiffness $e$
continuously increases with time due to the depletion of the ATP
inside the RBC  \cite{nirsam}. Correspondingly, the spinodal and
equilibrium areas (\eq{spineq}) decrease with time, and so does the
energy barrier that keeps the cell in a stressed state. The buckling
rate should eventually adjust to the rate of cytoskeleton
stiffening, as the pinching-off of the buckled membrane reduces the
RBC area and stabilizes the flat state until further stiffening.
Blebbing occurs when a thermal fluctuation moves the system over the
energy barrier. This requires a waiting time of order $\tau_0e^{\Delta F/k_{B}T}$},  where
$\tau_0\sim0.1 ms$ is the typical timescale for membrane
fluctuations on the length-scales of a single cytoskeleton unit\cite{seifert}, and
$\Delta F$ is the area-dependent energy barrier (\fig{full_buck}).
Accordingly, the RBC area at a given time is such that $\Delta
F=\kT\log{\tau_e/\tau_0}$, where $\tau_e$ is the characteristic time
of cytoskeleton stiffening.

\begin{figure}[h]
\centerline {\includegraphics[width=8cm]{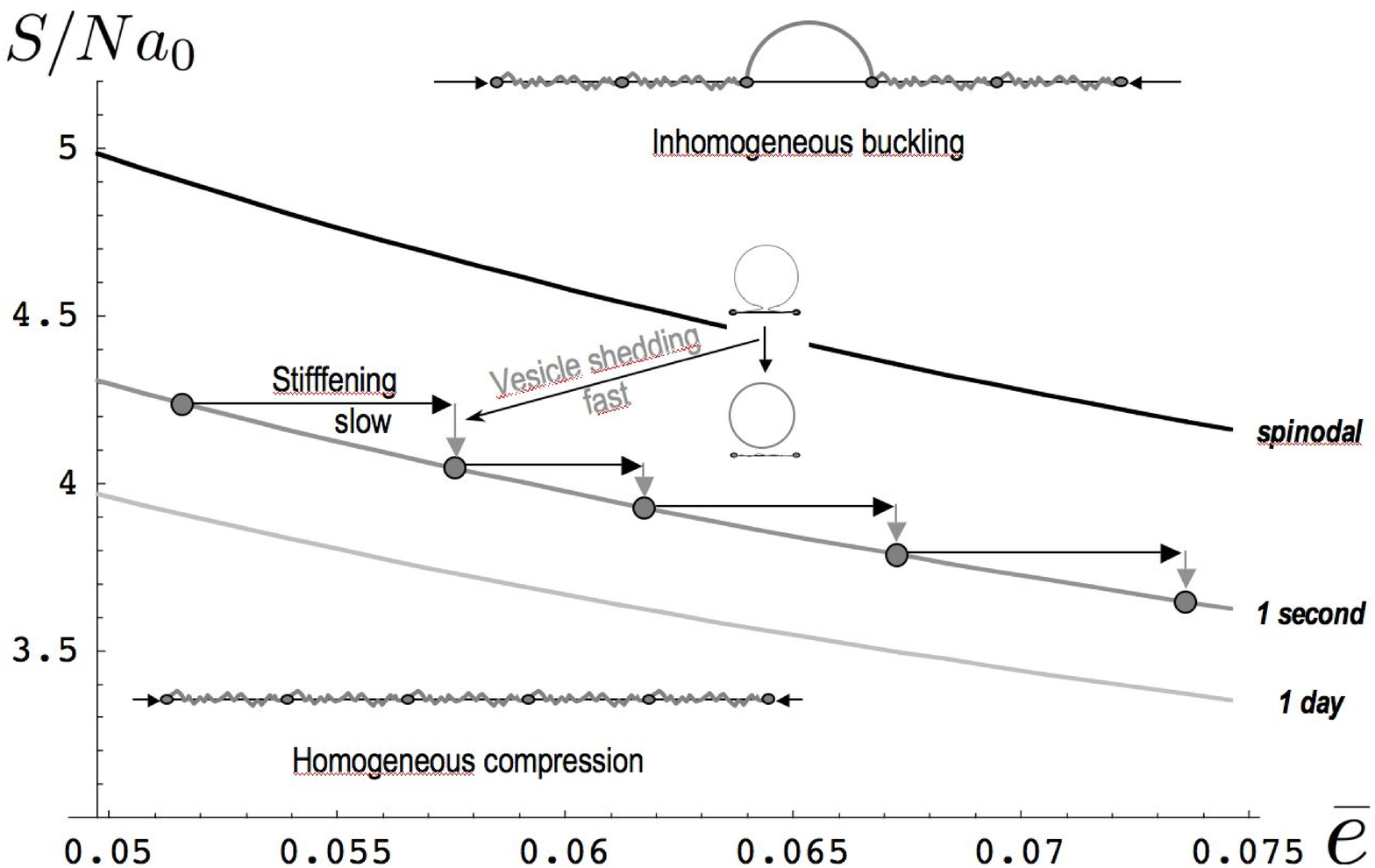}}
\caption{\label{aging} \small ``Non-equilibrium Phase diagram'' for RBC buckling. Cells with large membrane area (y-axis) are unstable to buckling. The critical area for buckling depends upon the (time-dependent) cytoskeleton stiffness (x-axis), and the rate at which the cytoskeleton stiffens. Aging RBCs follow trajectories showing cytoskeleton stiffening and the loss of membrane area (through the shedding of vesicles, represented as vertical arrows). Trajectories are such that the rate of vesicle shedding matches the rate of cytoskeleton stiffening ($\tau_e=\tau_0e^{\Delta F[S,\bar e]}$ see text). Three rates are represented ($\tau_e=1$day, $1$ second, and the spinodal line, \eq{spineq}, corresponds to infinitely fast stiffening). Parameters are the same as Figs.1-2}
\end{figure}

\fig{aging} shows the evolution of the cell membrane area for
various rates of cytoskeleton stiffening. Buckling occurs close to
$S=S|_{spin}$ in the limit of a very fast ATP depletion rate, and at
$S=S|_{eq}$ for unrealistically slow rate (not shown). Near the
spinodal line, an increase $\delta e$ of the cytoskeleton stiffness
leads to a decrease $\delta S=-S\delta e/(2e)$ of membrane area
(\eq{spineq}). The observed rate of membrane shedding from an aging
RBC is $20\%$ in $\sim100$ days, which translates to about one
$100$nm-sized vesicle an hour, consistent with an energy barrier at the transition $\Delta F\simeq 15\kT$. We predict this loss
of area to correlate with a $40\%$ stiffening of the spectrin
cytoskeleton (\fig{aging}), in agreement with mechanical and
structural measurements  \cite{fricke,sutera,ostafin}. The calculated metastable excess membrane area $S/Na_0\simeq 4$ (\fig{aging}) also agrees well with experimental observation \cite{atprafi}.

We conclude that the compression forces produced by an elastic network on an
attached fluid membrane may lead to buckling, and eventual
vesiculation of the membrane. These conditions occur under slow RBC
aging, during ATP depletion or Ca$^{2+}$ loading, which all
result in cytoskeleton stiffening  \cite{nirsam}. Note that
Ca$^{2+}$ loading occurs regularly at the neural synapse, where
internal vesicles respond by fusing and releasing their content
 \cite{kiss}. Our work shows that a cortical network may act to mechanically produces small vesicles, in a process entirely distinct from the classical Clathrin-based budding mechanism. It may therefore be relevant to some of the many biological systems containing
fluid membranes with an attached elastic network.

\begin{acknowledgments}
We thank the EU SoftComp NoE grant, the french ANR (P.S.) and the Robert Rees
Fund for Applied Research and the ISF (N.G.), for their support.
\end{acknowledgments}

\end{document}